\documentclass[trackchanges,twocolumn]{aastex701}
\usepackage{float}
\usepackage{booktabs}
\usepackage{graphicx}   % (caso use figuras)
\usepackage{amsmath}
\usepackage{newunicodechar}
\newunicodechar{−}{\ensuremath{-}}
\usepackage{xcolor}
\usepackage{makecell}
\usepackage{soul}

\begin{document}

\title{An extreme case of X-ray reflection in the symbiotic V648\,Car}

\author[]{Jhérssica Freitas}
\affiliation{Departamento de F\'isica, Universidade Federal de Sergipe, Av. Marechal Rondon, S/N, 49100-100, S\~ao Crist\'ov\~ao, SE, Brazil}
\email{jherssicafreitas@gmail.com}

\author[0000-0002-6211-7226]{Raimundo Lopes de Oliveira}
\affiliation{Departamento de F\'isica, Universidade Federal de Sergipe, Av. Marechal Rondon, S/N, 49100-100, S\~ao Crist\'ov\~ao, SE, Brazil}
\affiliation{Observat\'orio Nacional, Rua Gal. Jos\'e Cristino 77, 20921-400, Rio~de~Janeiro, RJ, Brazil}
\email{raimundo.lopes@academico.ufs.br}

\author[0000-0002-8286-8094]{Koji Mukai}
\affiliation{CRESST II and X-ray Astrophysics Laboratory, NASA/GSFC, Greenbelt, MD 20771, USA}
\affiliation{Department of Physics, University of Maryland, Baltimore County, 1000 Hilltop Circle, Baltimore, MD 21250, USA}
\email{Koji.Mukai@nasa.gov}

\begin{abstract}

V648 Car is a $\delta$-type symbiotic star known for its strong hard thermal X-ray emission, heavy local photoelectric absorption, and stochastic variability. We analyze \textit{NuSTAR} observations, complemented by nearly contemporaneous \textit{Swift}/XRT observations, to check for the expected presence and physical implications of Compton reflection in its X-ray spectrum. The spectra were fitted with X-ray models that accounted for thermal plasma components, different prescriptions to quantify the intrinsic absorption, the presence of the fluorescent iron line at 6.4\,keV, and a reflection component. Our results reveal a strong Compton reflection contribution, accounting for approximately 37\% of the total unabsorbed flux in the 3--50 keV band. The inclusion of reflection significantly improves the spectral fits and reduces the value for the maximum plasma temperature by a factor of 2 (from about 49 keV to 25 keV). If the maximum temperature is attributed to strong shock of a Keplerian flow, the inferred white dwarf mass decreases from about 20\% (from $1.2$ to $0.95\,M_{\odot}$), demonstrating that neglecting reflection can lead to significant overestimates of this parameter. Reflection models yield high reflection scaling factors ($R \sim 2$), inconsistent with a simple picture of a primary X-ray source above a reflecting disk. The timing analysis reveals stochastic flickering variability with no evidence for coherent periodic modulations, thus the accreting object is likely a non-magnetic white dwarf. We propose a multiple-reflection model to explain the extraordinary strength of reflection features in V648~Car.

\end{abstract}


\section{Introduction}
\label{sec:introduction}

Symbiotic stars are interacting binaries in which a compact object, typically a white dwarf (WD), accretes mass from the stellar wind of a red giant \citep{Kenyon1986, Mikolajewska2012, Munari2019}. V648~Car (SS73~17; CD$-$57~3057) was first identified by \citet{sanduleak_1973} and later characterized in greater detail through optical studies by \citet{pereira_2003}. The system subsequently drew interest in high-energy astrophysics when the BAT/Swift survey detected it as the hard X-ray source SWIFT J1010.1−5747 \citep{Tueller2005a,Tueller2005b}. This high-energy emission was independently confirmed by the IBIS instrument aboard INTEGRAL \citep{revnivtsev_2006,bird_2007}, prompting \citet{masetti_2006} to identify V648~Car as the definitive optical counterpart. Following these discoveries, \citet{Smith_2008} expanded the source's X-ray profile using observations from the Suzaku satellite. Swift observations were explored by \citet{Kennea_2009} and Chandra observations by \citet{Eze_2010}. Altogether, the X-ray observations revealed V648~Car to be a hard thermal ($kT_{\rm max} \sim 37$ keV) and variable X-ray source, with a strong and complex local absorption ($N_{\rm H} \sim 10^{23}$ cm$^{-2}$) and displaying a strong Fe K complex emission. It is consistent with the ``type $\delta$'' classification proposed for symbiotic systems by \citet{Luna2013}, to objects exhibiting hard, strongly absorbed X-ray emission.
Based on Gaia measurements, the system is located at a distance of $D = 1261^{+54}_{-47}$ pc \citep{Bailer_Jones_2021}.

A fraction of the hard X-ray photons from the primary emission of a ``type $\delta$'' symbiotic may undergo Compton scattering by relatively cold and dense material in the surroundings -- such as the WD surface or the inner regions of the accretion disk. This process, known as Compton reflection, produces a Compton hump (an excess of emission in the $\sim$10--30\,keV range) and is also accompanied by fluorescent emission seen as the Fe K$\alpha$ line at 6.4\,keV \citep{MagdziarzZdziarski1995}. Properly accounting for this component is essential, since its omission may lead to biased estimates of important physical parameters, such as the plasma temperature and, consequently, the white dwarf mass ($M_{\mathrm{WD}}$). However, despite its relevance, Compton reflection is still poorly explored in hard X-ray emitting symbiotic stars, being consistently identified in only a limited number of systems. Herein lies the core motivation for this work: V648~Car emerges as a well-suited target for this investigation, aimed at deepening our understanding of the reflection effect in symbiotic stars. 

In this work, we analyze NuSTAR observations of V648~Car to search for and characterize the Compton reflection component. The results are compared with the Swift/XRT observations that, as it is usual, offer a snapshot during the NuSTAR visits. We investigate its impact on the derived spectral parameters, with particular emphasis on the plasma temperature and the inferred $M_{\mathrm{WD}}$. This study provides the first detailed assessment of Compton reflection in this source and offers new perspectives on the accretion geometry and physical properties of $\delta$-type symbiotic stars.

\section{X-ray observations}

\subsection{NuSTAR}

V648~Car was observed by \textit{NuSTAR} (\textit{ObsID} 30201024002) for an effective exposure of 115~ks, spanning a total duration of 206~ks starting on 2017-01-08 at 19:56:12 UTC. Data from the FPMA and FPMB modules were processed using the HEASoft package (version 6.34) with calibration files version 20241126 available in the CALDB database. Event files were generated with the \texttt{nupipeline} task, adopting the standard filtering criteria. Source spectra were extracted from a circular region with a radius of 30 arcsec centered on the nominal source position, while background spectra were extracted from nearby source-free regions on the same detector. Spectral products, including redistribution matrix and ancillary response files, were generated using the \texttt{nuproducts} task. The NuSTAR spectra (FPMA and FPMB) were analyzed in the 3--50 keV energy range and grouped to a minimum of 25 counts per bin, allowing the use of $\chi^2$ statistic. 

\subsection{Swift}

Complementary X-ray observations were obtained with the X-ray Telescope (XRT) of the Neil Gehrels Swift Observatory under \textit{ObsIDs} 00081889002 and 00081889003, with exposures of approximately 2 ks each. The data were processed with the \texttt{xrtpipeline} task (v0.13.7) in HEASoft (v6.34), using 
calibration files version 20240522 in the CALDB database. Source spectra were extracted from circular regions with a radius of 45 arcsec, and background spectra were obtained from nearby source-free areas. Auxiliary response files were generated using \texttt{xrtmkarf}, while the response matrices were obtained directly from the CALDB for the \textit{Photon Counting} (PC) observation mode.

For the joint spectral analysis with the XRT/Swift data (0.3--10 keV), two statistical approaches were adopted: ungrouped spectra fitted with Cash statistics (C-stat) and grouped spectra (minimum of 25 counts per bin) analyzed with the $\chi^2$ statistic. The consistency between both methods was used to evaluate the robustness of the derived parameters.

\section{X-ray Spectral Modeling}
\label{sct:spctmodel}

The spectra of V648~Car obtained with \textit{NuSTAR} (FPMA and FPMB) were fitted simultaneously using the \texttt{XSPEC} software version 12.14.1 \citep{Arnaud1996}. Solar abundances from \citet{Wilms2000} were adopted. Interstellar absorption was accounted for with the \texttt{tbabs} component, that takes into account the contribution of the gas-phase, the grain-phase, and the molecules in the ISM, and the effects of the local photo-electric absorption. More complex intrinsic absorption models, such as partial covering with \texttt{pcfabs} and continuous column density distributions with \texttt{pwab}, were also tested. The \texttt{pcfabs} includes a dimensionless covering fraction parameter for a given absorption column. As for the \texttt{pwab}, the parameter are the lower and upper limits for the equivalent in Hydrogen column ($N_{\rm H,min}$ and $N_{\rm H,max}$), and the index ($\beta$) of a power-law distribution of covering fraction as a function of column density. The $N_{\rm H,min}$ in \texttt{pwab} was fixed to 10$^{15}$\,cm$^{-2}$ as it is not constrained from the available spectra. The thermal emission was described using the single optically thin plasma model, \texttt{apec}, and its combination in the \texttt{mkcflow} (\textsc{switch}\,=\,2) to emulate a cooling flow, while the fluorescent Fe K$\alpha$ line was modeled with a Gaussian (\texttt{gauss}) component centered (and fixed) at 6.4\,keV, with $\sigma$ fixed to 10$^{-3}$ keV. The required redshift parameter in \texttt{mkcflow} model was fixed to 2.944$\times$10$^{-7}$, following GAIA distance and $H_0$ = 70\,km\,s$^{-1}$\,Mpc$^{-1}$. The low temperature component of \texttt{mkcflow} ($kT_{\rm low}$) was set to 0.0808\,keV  in all cases, the lowest possible value in the model, as it is not constrained from the available spectral coverage. The \texttt{reflect} model was added to account for the expected Compton reflection, assuming that it occurs on neutral material \citep{1995MNRAS.273..837M}. As a convolution model, the \texttt{reflect} modifies an input continuum from a primary X-ray source that, in our case, is described by a cooling flow model (or a single plasma as a first try). The geometry and physical state of the reflector are parametrized with the \texttt{rel$_{\rm refl}$} and \texttt{cos$\theta$} parameters. 

The \texttt{rel$_{\rm refl}$}, or reflection amplitude, is such that the value of 1 corresponds to an isotropic source illuminating an infinite, plane-parallel slab, where half of the primary photons reach the observer directly, and the other half are intercepted by the slab. In other words, it quantifies the relative intensity of the reflected component. The \texttt{cosIncl} is the cosine of the inclination angle ($\theta$) between the observer's line of sight and the reflecting slab's surface normal, which need not necessarily correspond to that of the orbital plane. In this way, \texttt{cos$\theta$} equal to 1 or 0 corresponds to face-on and edge-on geometries, respectively, so that it  quantifies the projected optical depth and path length of the scattered photons.
The spectral response was extended to 100\,keV (\texttt{xset REFLECT\_MAX\_E} = 100.0).

\section{Results}

\subsection{Spectral analysis}

We performed a systematic spectral analysis by testing progressively more complex scenarios, evolving from a simple single-temperature plasma to multi-temperature models that account for reflection and complex absorption. Initially, a single-temperature plasma model (\texttt{tbabs*(apec+gauss)}; M1) failed to provide an acceptable description of the data, leaving systematic residuals in the ionized Fe\,K complex (6.7--6.9 keV) and yielding an over-solar metal abundance. Transitioning to a multi-temperature cooling-flow plasma model (\texttt{tbabs*(mkcflow+gauss)}; M2) significantly improved the fit quality and lowered the inferred metal abundance, though a noticeable systematic excess remained at higher energies (20--30 keV). To address this high-energy excess, we incorporated a Compton reflection component. While adding reflection to a single-temperature plasma (\texttt{tbabs*(reflect*apec+gauss)}; M3) yielded only marginal improvements, the best description among the standard emission scenarios was obtained by combining reflection with the multi-temperature plasma (\texttt{tbabs*(reflect*mkcflow+gauss)}; M4). This combination successfully accounted for the hard X-ray excess, decreasing the maximum plasma temperature and returning near-solar metallicity. In all reflection models, the inclination angle parameter ($\theta$) was fixed at $\cos\theta = 0.45$ (see Section \ref{sct:dependcos}).

Given the heavily obscured nature of the source, we further explored complex absorption geometries by building upon the M4 framework. We tested a partial covering scenario (\texttt{tbabs*pcfabs*(reflect*mkcflow+gauss)}; M5) and a continuous distribution absorption model (\texttt{tbabs*pwab*(reflect*mkcflow+gauss)}; M6). The partial covering model implies that the vast majority of the X-ray emitting region is obscured by dense local material. Alternatively, the power-law index $\beta$ obtained in M6 suggests a broad distribution of absorbing column densities along the line of sight. For both models, the Galactic foreground column density was fixed based on the HI4PI survey \citep{2016A&A...594A.116H} via the 3DN$_{\rm H}$-tool. Notably, despite these different geometric prescriptions for the intrinsic absorption, both complex models converged on remarkably consistent physical properties for the underlying source. Furthermore, the equivalent width of the Fe K$\alpha$ line remained stable at $\rm EW \sim 0.33$ keV across all tested scenarios.

\begin{figure}[t]
\centering
\includegraphics[width=\columnwidth]{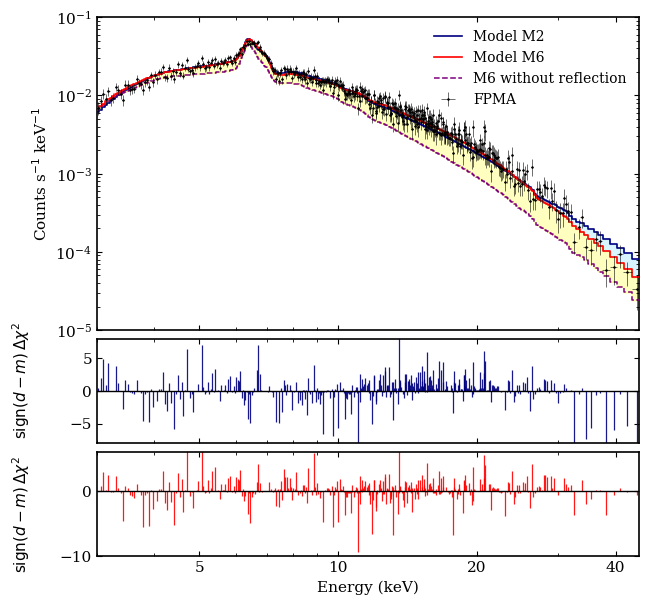}
\caption{The upper panel shows the observed spectrum together with the best-fit model without (M2) and with reflection (M6), and the latter (M6) after removing the reflection component by hand; see legend. The middle and lower panels present the residuals for models M2 and M6, respectively.}
\label{fig:comparison_reflection}
\end{figure}

The best-fit parameters, including metallicity and mass accretion rates for all models, are summarized in Table~\ref{tab:reflection_models}. Figure~\ref{fig:comparison_reflection} highlights the critical role of the reflection component by comparing the pure multi-temperature fit (M2) against the complex absorption and reflection model (M6). The substantial flux exces attributed to Compton reflection is visually isolated in Fig. \ref{fig:comparison_reflection} (yellow-hatched area) by manually setting the reflection component of M6 to zero post-fit.

\begin{table*}
\centering
\caption{Best-fit parameters from the models applied to the \textit{NuSTAR} spectra.}
\label{tab:reflection_models}
\setlength{\tabcolsep}{3.5pt} % Espaçamento confortável entre as colunas
\begin{tabular}{lccccccccc}
\toprule
Model & \texttt{tbabs} & \multicolumn{3}{c}{\texttt{pcfabs} / \texttt{pwab}} & \texttt{apec}/\texttt{mkcflow} & \texttt{reflect} & & & \\ 
\cline{3-5}
 & $N_{\mathrm{H, gal}}$ & $N_{\mathrm{H, int}}$ & CF & $\beta$ & $k T$/$k T_{\mathrm{high}}$ & $R$ & $Z$ & $\dot{M}$ & $\chi^2_{\rm red}$/dof \\
 & \multicolumn{2}{c}{($10^{22}$ cm$^{-2}$)} & & & (keV) & & ($Z_{\odot}$) & & \\
\midrule

\parbox{5.2cm}{\raggedright M1: {\tt tbabs*(apec+gauss)}} & 
$13.33^{+0.75}_{-0.76}$ & ... & ... & ... & $22.58^{+0.94}_{-0.81}$ & ... & $2.06^{+0.31}_{-0.27}$ & ... & 1.30/870 \\ \addlinespace

\parbox{5.2cm}{\raggedright M2: {\texttt{tbabs*(mkcflow+gauss)}}} & 
$16.30^{+0.87}_{-0.83}$ & ... & ... & ... & $48.73^{+2.85}_{-2.82}$ & ... & $1.47^{+0.20}_{-0.18}$ & $\sim 0.91$ & 1.14/869 \\ \addlinespace

\parbox{5.2cm}{\raggedright M3: {\texttt{tbabs*(reflect*apec+gauss)}}} & 
$11.16^{+0.83}_{-0.82}$ & ... & ... & ... & $15.10^{+0.82}_{-0.59}$ & $1.86^{+0.34}_{-0.36}$ & $1.34^{+0.14}_{-0.12}$ & ... & 1.25/870 \\ \addlinespace

\parbox{5.2cm}{\raggedright M4: {\texttt{tbabs*(reflect*mkcflow+gauss)}}} & 
$13.38^{+0.97}_{-0.99}$ & ... & ... & ... & $25.29^{+1.64}_{-1.63}$ & $2.37^{+0.51}_{-0.41}$ & $0.96^{+0.08}_{-0.08}$ & $\sim 1.18$ & 1.03/869 \\ \addlinespace

\parbox{5.2cm}{\raggedright M5: {\texttt{tbabs*pcfabs*\\(reflect*mkcflow+gauss)}}} & 
$1.37^{(*)}$ & $16.54^{+0.96}_{-0.95}$ & $0.86^{(*)}$ & ... & $25.52^{+1.00}_{-0.97}$ & $2.12^{+0.28}_{-0.25}$ & $0.92^{+0.05}_{-0.05}$ & $\sim 1.23$ & 1.03/869 \\ \addlinespace

\parbox{5.2cm}{\raggedright M6: {\texttt{tbabs*pwab*\\(reflect*mkcflow+gauss)}}} & 
$1.37^{(*)}$ & $10.64^{+0.60}_{-0.95}$ & ... & $0.96^{+0.17}_{-0.16}$ & $25.27^{+0.97}_{-0.94}$ & $2.35^{+0.27}_{-0.24}$ & $0.93^{+0.05}_{-0.05}$ & $\sim 0.79$ & 1.03/869 \\

\bottomrule
\end{tabular}
\begin{flushleft}
\small $\dot{M}$ in units of $10^{-9}\,M_\odot\,\mathrm{yr}^{-1}$. CF denotes the covering fraction of the \texttt{pcfabs} model, while $\beta$ is the power-law index describing the column-density distribution in the \texttt{pwab} model. (*) Parameter fixed during fitting (see Section \ref{sct:spctmodel}).
\end{flushleft}
\end{table*}

%%%%

The addition of the short XRT/Swift exposure ($\sim$\,2\,ks) to the joint spectral analysis did not further constrain the fit. However, we verified that the best-fit parameters derived from the NuSTAR spectra alone remain fully consistent with the XRT/Swift data.

\subsection{Timing investigation}
\label{sct:timing}

In this section, we analyze the timing properties of V648~Car to characterize its stochastic behavior and search for coherent periodic signals. Such modulations could arise from binary orbital effects or much shorter-period pulsations, which are observed in some non-magnetic CVs and originate within the boundary layer.

\begin{figure*}[t]
\centering
\includegraphics[width=\textwidth]{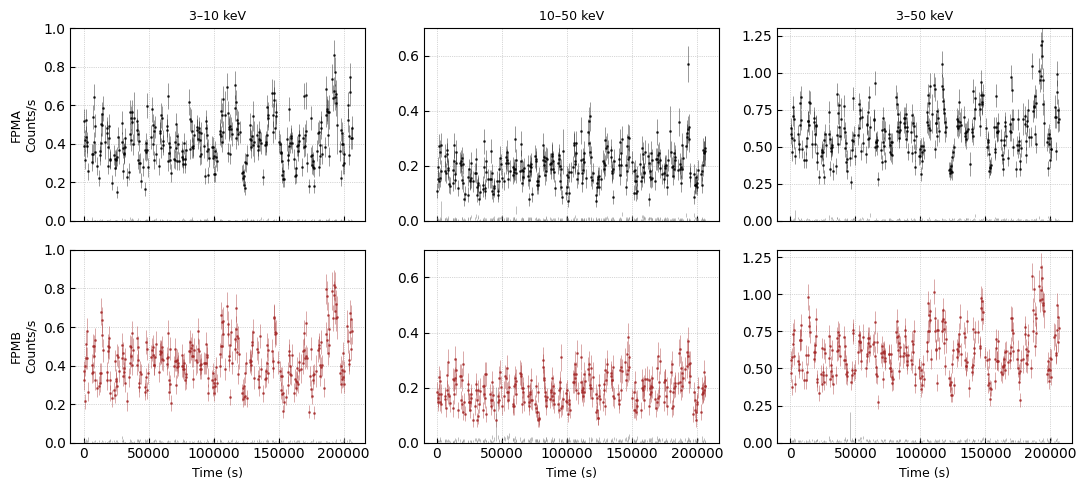}
\caption{
Light curves of V648~Car 
in the 3--10\,keV, 10--50\,keV, and 3--50\,keV bands
for the FPMA (upper) and FPMB/NuSTAR (bottom) detectors binned to 400 s. The gray ``grass'' at the zero level in each panel corresponds to the background contribution.}
\label{fig:lightcurves}
\end{figure*}

The X-ray variability of V648~Car was investigated using the \textit{NuSTAR} light curves extracted in the 3--10\,keV, 10--50\,keV, and 3--50\,keV energy bands, analyzed independently for the FPMA and FPMB instruments. The light curves, generated with a time binning of 400\,s for clarity, are presented in Fig.~\ref{fig:lightcurves}.

Light curves from both detectors are mutually consistent. They exhibit significant variability throughout the observation, with irregular fluctuations on timescales of a few ks. There is no visual evidence of coherent periodic modulation, nor, as discussed below, in the periodogram. In the 3--50\,keV band, the light curves show average count rates of $0.59$\,counts\,s$^{-1}$ for FPMA and $0.61$\,counts\,s$^{-1}$ for FPMB, with fractional RMS variability of approximately 30\% and 26\%, respectively, indicating significant intrinsic variability during the entire observation. This behavior is characteristic of stochastic \textit{flickering} variability, which is frequently associated with disk-fed accretion. This includes the \textit{boundary layer} (BL) region, where fluctuations driven by local instabilities and variations in the accretion rate appear to be responsible for the rapid changes in the observed flux.

\begin{figure}[t]
\centering
\includegraphics[width=\columnwidth]{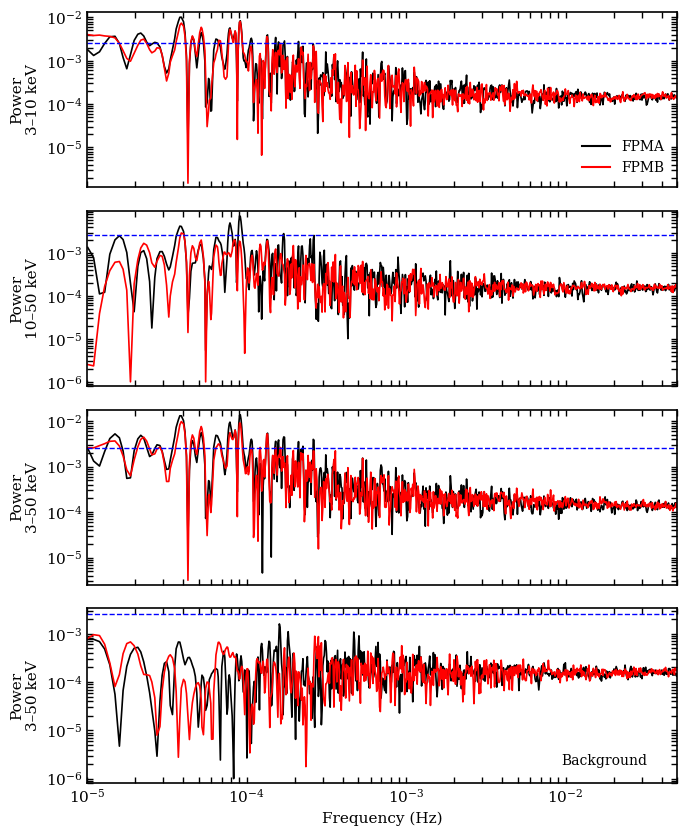}
\caption{Periodograms from V648~Car light curves in the 3--10\,keV, 10--50\,keV, and 3--50\,keV bands (from top to bottom), with the background periodogram for the 3--50\,keV band shown in the bottom panel (FPMA and FPMB in black and red, respectively). The blue dashed line indicates the power level corresponding to a false-alarm probability of 1\%.}
\label{fig:periodogram}
\end{figure}

The search for periodic modulations in X-rays was performed using the Lomb--Scargle periodogram, implemented in the \texttt{astropy} package \citep{Lomb1976,Scargle1982,VanderPlas2012}. It was applied to light curves binned to 10\,s in the 3--10 keV, 10--50 keV, and 3--50 keV bands from the Nyquist limit of 20\,s up to one half of the total coverage, so $\sim$\,100\,ks.
We also applied the periodogram to light curves from events in background regions to distinguish possible instrumental signals from those associated with the source's intrinsic variability. The results are shown in Fig.~\ref{fig:periodogram}, where the blue dashed line indicates the power level corresponding to a False Alarm Probability (FAP) of 0.01 (formally, 1\% chance that noise could create a peak this high).

The periodograms of V648~Car exhibit multiple power peaks distributed above $\sim\,5\times 10^{3}$ s, with the strongest ones corresponding to periods of approximately 10.9 ks and 25.6 ks, all formally above the exceeding threshold of FAP\,=\,0.01 from both FPMA and FPMB light curves. %However, red noise is clearly visible from the periodogram, as it is usual in accreting white dwarfs following the stochastic pattern in brightness fluctuation, and the observed peaks...
However, the interpretation on them requires caution as FAP is calculated under the assumption of white noise and may therefore overestimate the significance of low-frequency features in the presence of red-noise variability. As pointed out by \citet{bruch2022}, red noise manifests as an excess of power toward low frequencies and a progressive decline of power with increasing frequency, such as is the case for V648~Car. Such a stochastic variability can generate prominent peaks superimposed on a red-noise continuum without necessarily implying the presence of a coherent periodic modulation \citep[see][]{dobrotka_2025}. 
%Therefore, although several peaks formally exceed the FAP = 1\% level, 
After all, we understand that the distribution of multiple low-frequency peaks ($>$ 5\,times 10$^{3}$ s), rather than being dominated by a single frequency, suggests that these structures are more likely associated with stochastic accretion-driven variability instead of intrinsic periodic modulations in V648 Car.
%on the timescales accessible with the NuSTAR observations (from the Nyquist limit of 20\,s up to one half of the total coverage, so $\sim$\,100\,ks).
It is also important to take into account that NuSTAR is in a low Earth orbit, meaning that the Earth occults the target every $\sim 97$ minutes ($\sim$ 5800 s), introducing periodic gaps into the data. Consequently, periodicities close to 5800 s and its sub-harmonics/harmonics (e.g., 2900 s or 11600 s) are extremely difficult to verify, as the satellite's orbital motion introduces a strong artificial signal at these frequencies.

\section{Discussion}

V648~Car is a luminous, strongly absorbed, and hard thermal X-ray source. The total unabsorbed flux in the 3–50~keV band is $\sim$ 3.9 $\times$ 10$^{-11}$ erg\,cm$^{-2}$\,s$^{-1}$ (derived from model M6), which, at the source distance, corresponds to a luminosity of $\sim$ 7.4 $\times$ 10$^{33}$ erg\,s$^{-1}$. Our timing analysis did not reveal any periodicity attributable to the spin of a magnetic WD, so we adopt a non-magnetic WD model as our working hypothesis in interpreting our results. The most notable feature of V648~Car, as already noted by \cite{Smith_2008} and by \cite{Eze_2010}, is the strong reflection features (Compton hump and Fe K$\alpha$ line). Therefore, we concentrate our discussion on the observational confirmation and possible interpretations of the strong reflection signatures.

\subsection{Confirmation of reflection and its impact on the shock temperature}
%\subsection{Reflection, luminosity, and physical implications}

Our spectral fits to the NuSTAR data have confirmed that reflection features are strong in V648~Car. In our fits, Compton reflection contributes significantly to the observed X-ray emission of this symbiotic star, accounting for about 37\% ($\sim$\,1.43$\times$ 10$^{-11}$~erg~cm$^{-2}$~s$^{-1}$) of the total emission (see the yellow-hatched area in Fig. \ref{fig:comparison_reflection}). This means that a substantial fraction of the radiation emitted by the hot plasma is reprocessed by cold, optically thick material in the vicinity of the WD and subsequently re-emitted toward the observer. The need for including this component is clearly demonstrated by comparing models with and without reflection (Fig. \ref{fig:comparison_reflection}). In the absence of reflection, a systematic excess is observed in the 20--30~keV range that cannot be reproduced by purely thermal models, even when a continuous temperature distribution and complex absorption are considered. This excess, characteristic of the so-called \textit{Compton hump}, is fully accounted for once reflection is included. As shown in Fig.~\ref{fig:comparison_reflection}, the model without reflection overestimates the hard X-ray flux ($>$\,30\,keV; the blue-hatched area), while the inclusion of reflection properly reproduces the high-energy excess and significantly improves the spectral fit (red line).

The derived reflection scaling factors from models that consider a cooling flow as the primary source of X-rays, in the range $R \sim 2.1$--$2.4$, are significantly greater than unity. As discussed in Section \ref{sct:dependcos}, this result remains robust even when \texttt{cos$\theta$} is allowed to vary freely. Furthermore, we argue in Section \ref{sct:geometry} that such a high scaling factor is a consequence of the geometry of emission and reflection sites in V648~Car being more complex than what is assumed by the \texttt{reflect} model, and does not imply a problem with the reflection interpretation in general. In the remainder of this section, we discuss how the estimated plasma temperature changes with the introduction of the reflection component and the subsequent implications for the WD mass.

The inability of purely thermal models to reproduce the high-energy spectrum has direct consequences for determining the system’s physical parameters. In the absence of reflection, the spectral fit compensates for the excess hard X-ray flux by artificially increasing the maximum plasma temperature. This is already true for the unrealistic single plasma model (single \texttt{apec}), with the inferred temperature dropping from $\sim$23 to $\sim$15 keV, but even more so for the cooling flow model. The \texttt{mkcflow} yielded $kT_{\mathrm{max}} \sim 49$~keV without reflection, and $\sim$ 25\,keV if reflection is included, so by a factor 2. This indicates that part of the emission previously attributed to the primary X-ray emitter, the hot plasma, is actually produced by reprocessed radiation from the primary source in the cold material near its site.

Assuming that the hard X-ray emission originates in the optically thin portion of the boundary layer, the maximum shock temperature derived from the \texttt{mkcflow} model can be used to estimate  $M_{\mathrm{WD}}$. Following the classical treatment of accretion in non-magnetic systems, approximately half of the gravitational potential energy is expected to be released in the accretion disk, while the remaining half is dissipated in the boundary layer where the Keplerian flow undergoes a strong shock. Under this assumption, we calculated $M_{\mathrm{WD}}$ using the mass--radius relation for cold WDs proposed by \citet{Pringle1975}, as applied by \citet{Byckling2010} to dwarf novae and symbiotic systems.

Without including reflection, the higher temperature obtained from model M2 implies  $\sim$\,$M_{\mathrm{WD}} \sim 1.2\,M_{\odot}$. On the other hand, when reflection is taken into account, the lower temperature derived from models M4--M6 leads to a significantly smaller mass estimate of $M_{\mathrm{WD}} \sim 0.95\,M_{\odot}$. This corresponds to a reduction of about 20--25\%, demonstrating that neglecting reflection may lead to systematic overestimates of $M_{\mathrm{WD}}$. Similar behavior has been reported for other hard X-ray symbiotic stars, such as RT~Cru and SU~Lyn, for which the inclusion of reflection also resulted in lower plasma temperatures and, consequently, lower $M_{\mathrm{WD}}$ estimates \citep{2018A&A...616A..53L,Lopes2018}.

In reality, the boundary layer plasma does not appear to reach the maximum possible shock temperature for a Keplerian flow. In a sample of quiescent dwarf novae, the observed $kT_{\mathrm{max}}$ was about 60\% of the ``strong shock from Keplerian'' case \citep{Mukai2022}. If we apply this purely empirical correction, the values are $M_{\mathrm{WD}} \sim 1.16\,M_{\odot}$ for $kT_{\mathrm{max}} \sim 25 $~keV with reflection (maximum possible shock temperature of $\sim 42$~keV) and $M_{\mathrm{WD}} \sim 1.345\,M_{\odot}$ for $kT_{\mathrm{max}} \sim 46 $~keV without.

\subsection{Dependence of the reflection factor on the viewing angle}
\label{sct:dependcos}

In the spectral fits that include the reflection component (M3, M4, M5, and M6), the geometrical parameter associated with the \texttt{reflect} model, defined as the cosine of the view angle $\theta$ between the observer's line of sight and the normal to the reflecting surface, was initially fixed at the intermediate value of \texttt{cos$\theta$} = 0.45. This choice was adopted because, when allowed to vary freely, the fit exhibited strong degeneracy, preventing reliable determination of the uncertainties of \texttt{cos$\theta$} and yielding statistically unstable and physically  inconsistent solutions for the remaining spectral parameters.

\begin{figure}[t]
\centering
\includegraphics[width=\columnwidth]{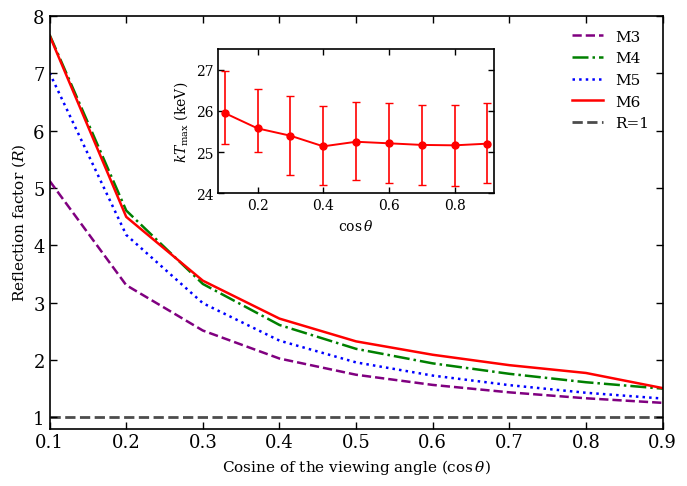}
\caption{The reflection factor (for M3 to M6) and $kT_{\max}$ (for M6) as a function of the cosine of the inclination angle between the observer's line of sight and the normal of the reflecting surface.}
\label{fig:cosincl}
\end{figure}

To evaluate the dependence of the reflection factor on the assumed geometry, we performed tests by systematically varying the parameter \texttt{cos$\theta$}. We find that the reflection factor decreases as this value increases, that is, as the reflecting surface is viewed in a more \textit{face-on} configuration. This behavior is consistent with the angular dependence expected for Compton reflection processes, in which the efficiency of back-scattered radiation increases for more inclined geometries \citep{MagdziarzZdziarski1995}.
It is worth noting that the $kT_{\max}$ from \texttt{mkcflow} is consistent within the $1\sigma$ confidence level for a given model, regardless of the assumed value of \texttt{cos$\theta$} in the \texttt{reflect} component -- see the inset of Fig.~\ref{fig:cosincl} for M6.
Despite the trend of $R$ versus \texttt{cos$\theta$}, a particularly important result is that the values of $R$ remain systematically high accross the entire explored range of \texttt{cos$\theta$}. Even for nearly \textit{face-on} configurations, the models return $R > 1$, while for higher inclinations the values can significantly exceed unity (see Fig. \ref{fig:cosincl}). This behavior indicates that the strong reflection component observed in V648~Car is not a consequence of a particular inclination angle, but rather an intrinsic property of the system.

Furthermore, the similarity in the behavior of $R$ among the different spectral models, regardless of the treatment adopted for local absorption, suggests that the strength of the reflection component is only weakly sensitive to uncertainties associated with absorption modeling, reinforcing the idea that the reflected component plays a dominant role in shaping the hard X-ray spectrum.

\subsection{Geometrical and physical constraints} \label{sct:geometry}

The \texttt{reflect} model assumes a primary source located above a flat accreting surface that extends to infinity.  In this geometry, the reflecting medium covers exactly half the sky (a solid angle of $\Omega$\,=\,2$\pi$ steradians), which implicitly caps the reflection fraction $R$ at a maximum value of 1, by definition. The fact that we derive a reflection scaling factor of above 2 is a concern in the context of the geometry assumed in the \texttt{reflect} model (as already pointed out by \citealt{Eze_2010}). However, there are geometries that result in such strong reflection signatures. One possible analogy is the Seyfert~2 type active galactic nucleus (AGN) in which the primary continuum is completely hidden from our view by a Compton-thick absorber, and all we observe is the reflected emission.  Since the scaling factor of \texttt{reflect} model is the ratio of reflected to primary emission in the observed spectrum, an arbitrarily high value can result if we hide some fraction of the primary continuum.  A similar interpretation was offered for the anomalous hard X-ray faint state of the symbiotic star, CH~Cyg \citep{Mukai2007}. For this to work for V648~Car, there must be an absorber with substantially higher $N_{\mathrm{H}}$ than indicated by our partial covering absorber fit. While not impossible, this scenario might require fine-tuning in terms of the properties of intrinsic absorber.

\begin{figure}[t!]
\centering 
\includegraphics[width=\columnwidth, trim={6cm 4cm 3cm 10cm}, clip=true]{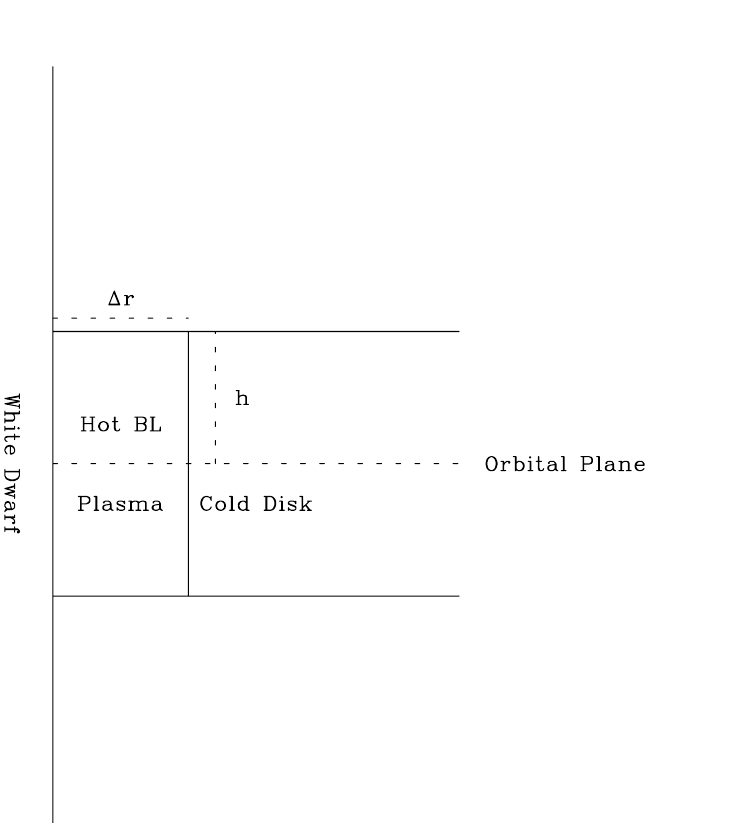}\caption{A schematic diagram of the proposed geometry.  In this scenario, hot plasma in the compact boundary layer (with radial extent $\Delta r$), sandwiched between the white dwarf surface and the Keplerian accretion disk of height $h$, is the site of the primary X-rays.  Both the white dwarf surface and the inner edge of the disk are reflection sites.}
\label{fig:blrefl}
\end{figure}

Another potential explanation for the high reflection factor is multiple reflection. In the case of AGN, this can arise from clumpy or corrugated disk \citep{Ross2002} or from gravitational light bending \citep{Wilkins2020}, both of which are known to enhance the 6.4 keV line and the Compton hump.
Here, we envision a scenario in which the X-ray emitting boundary layer is sandwiched between the accretion disk and the WD surface. The temperatures of both components are sufficiently low (likely below 10$^5$ K) for Fe atoms to retain their K-shell electrons, thereby satisfying the ``cold reflection'' condition. A schematic cross-section of this geometry is illustrated in Fig. \ref{fig:blrefl}.  The incoming disk has a half-thickness $h$ and undergoes a shock at a distance $\Delta r$ from the WD surface. Note that while $h$ typically denotes the shock height in the literature of magnetic CVs, here $\Delta r$ plays an equivalent role.

If the cold disk is optically thick to Compton scattering, both its inner edge and the WD surface will serve as active reflection sites. Provided that $\Delta r$ is a small fraction of the WD radius, virtually all downward-emitted X-rays intersect the WD surface, independently producing a reflection fraction amplitude of $R$\,$\sim$\,1. Furthermore, if $\Delta r < h$, only a small fraction of the photons can escape the boundary layer without interacting with the WD surface or the inner disk, which can lead to a reflection fraction $>$1.  Moreover, a reflected photon has a high probability of interacting with another reflecting surface, similar to an emission region at the bottom of a well as envisioned by \cite{Ross2002}.  When a spectrum including photons that have experienced multiple reflections is fit with the \texttt{reflect} model in the usual manner, the best-fit value of the reflection fraction can plausibly exceed 2.0.

An in-depth investigation of this scenario requires a ray-tracing simulation using a good knowledge of the geometry, and is beyond the scope of this paper.  Instead, we have conducted a simple, proof-of-concept study of this situation by replacing \texttt{reflect*mkcflow} in model M6 with \texttt{reflect*reflect*mkcflow}. When the reflection fraction of one of the \texttt{reflect} model is fixed at 1.0, that of the other converged to a value of 0.79$^{+0.17}_{-0.14}$, while the \texttt{mkcflow} parameters remained virtually unchanged; when both were allowed to vary, we obtained a fit with reflection fractions of 0.82 and 0.96, although they are degenerate and fit failed to provide a reliable range. This simplified exercise suggests that the multiple reflection scenario is plausible, pending a more detailed investigation.

We therefore adopt this sandwich geometry with multiple reflection as the working hypothesis to explain the extraordinarily strong reflection features in V648~Car. In the next section, we consider why this scenario can apply to V648~Car, but not to other hard X-ray bright, non-magnetic, accreting WDs.

\subsection{V648~Car among hard X-ray bright non-magnetic systems} \label{sct:comparison}

Among the 5 symbiotic stars detected in the Swift/BAT hard X-ray survey \citep{Kennea_2009, 10.1093/mnrasl/slw087}, V648~Car stands out for having such strong reflection features.  We therefore compare these systems, except for CH~Cyg because it has been proposed to be magnetic \citep{Pichard_Marcano2026}.
We also consider the case of SS~Cyg, an X-ray bright dwarf nova, since it is the standard laboratory for studying the X-ray properties of the WD boundary layer. We present the X-ray properties of these systems below and provide a summary in Table \ref{tab:comparison}. 

\begin{table*}
\centering
\caption{Comparison between V648~Car and other hard X-ray bright non-magnetic accreting white dwarf systems. All parameters are reported from spectral models that include a Compton reflection component.}
\label{tab:comparison}
\begin{tabular}{lccccccc}
\hline
Source &
Observation &
$N_{\rm H}$ &
CF &
$kT_{\rm max}$ &
$R$ &
EW(Fe K$\alpha$) &
$L_{\rm X}$ \\
&
&
($10^{22}$ cm$^{-2}$) &
&
(keV) &
&
(eV) &
(erg s$^{-1}$) \\
\hline

RT Cru $^{1}$
& NuSTAR+Swift
& $12 \pm 5 (a)$
& $0.61 \pm 0.10$
& $53 \pm 4$
& $0.77 \pm 0.21$
& $230$--$350$
& $4.9 \times 10^{34}$
\\

SU Lyn $^{2}$
& NuSTAR
&  $22.5^{+2.4}_{-3.0}(b)$
& ---
& $16.3^{+4.2}_{-2.7}$
& $0.9^{+1.0}_{-0.8}$
& $150$--$380$
& $4.9 \times 10^{32}$
\\

T CrB $^{3}$
& NuSTAR+Swift
& $43 \pm 3 (a)$
& $0.93 \pm 0.06$
& $40 \pm 3$
& $1.1 \pm 0.2$
& $191 \pm 2$
& $4.5 \times 10^{33}$
\\

SS Cyg (quiescence) $^{4}$
& XMM+NuSTAR
& $0.00158 (b)$
& ---
& $22.8^{+1.5}_{-1.1}$
& $1.25^{+0.27}_{-0.20}$
& $\sim80$
& $9.3 \times 10^{31}$
\\

SS Cyg (outburst) $^{4}$
& XMM+NuSTAR
& $3.4^{+1.0}_{-0.9} (a)$
& $0.23 \pm 0.02$
& $8.4^{+0.3}_{-0.3}$
& $1.31^{+0.30}_{-0.30}$
& $\sim80$
& $3.3 \times 10^{33}$
\\

V648 Car
& NuSTAR
& $10.64^{+0.60}_{-0.95} (a)$
& $0.96$
& $25.27^{+0.97}_{-0.94}$
& $2.35^{+0.27}_{-0.24}$
& $330$
& $7.4 \times 10^{33}$
\\

\hline
\end{tabular}

\vspace{0.2cm}

\begin{minipage}{0.97\linewidth}
\footnotesize

\textbf{Notes.}
$^{a}$ Intrinsic absorption column density. For RT Cru, T CrB and SS Cyg during outburst, $N_{\rm H}$ corresponds to the partial-covering absorber. For V648~Car, $N_{\rm H}$ corresponds to the maximum column density of the \texttt{pwab} absorption model.

$^{b}$ Absorption column density associated with the \texttt{TBabs} component.

X-ray luminosities were derived in different energy bands in the original studies: RT Cru (0.3--80 keV), SU Lyn (3--30 keV), T CrB (0.3--50 keV), SS Cyg (0.01--100 keV bolometric), and V648~Car (3--50 keV).

\textbf{References.}
$^{1}$ \citet{2018A&A...616A..53L},
$^{2}$ \citet{Lopes2018},
$^{3}$ \citet{Luna_2019},
$^{4}$ \citet{2023ApJ...957...33D}

\end{minipage}

\end{table*}

Using data obtained with \textit{Suzaku}, \textit{NuSTAR}, and \textit{Swift}, \citet{2018A&A...616A..53L} showed that RT~Cru exhibited complex and variable local photoelectric absorption.
A Compton reflection hump was detected and associated with reflection with an amplitude of $0.77 \pm 0.21$.
The neutral $\text{Fe K}\alpha$ fluorescence line at $6.4 \text{ keV}$, with an equivalent width ranging from 230 eV to 350 eV, indicates structural changes in the reflecting matter. The maximum post-shock plasma temperature was inferred to be $kT$ = 53 $\pm$ 4 keV, corresponding to a WD mass of 1.25\,$\pm$\,0.02 M$_\odot$; omitting the reflection component overestimated the plasma temperature at 64\,$\pm$\,5 keV.

Similarly, SU~Lyn is a hard X-ray symbiotic with complex and variable local absorption, in which reflection was also identified \citep{10.1093/mnrasl/slw087,Lopes2018}. The reflection amplitude was determined to be near unity, with the primary plasma emission coming from a shock cooling down from $kT$\,=\,21.1$^{+1.9}_{-2.6}$\,keV for a WD having at least 1\,M$_{\odot}$ -- or 16.3$^{+4.2}_{-2.7}$\,keV without reflection.

The symbiotic recurrent nova T~CrB, as presented by \citet{Luna_2019} from Swift and NuSTAR observations, exhibited, during a rising phase of its optical brightening, a plasma with a maximum temperature of $kT = (40 \pm 4)$ keV under complex absorbing conditions, with a reflection amplitude of 1.1\,$\pm$\,0.2.

For the dwarf nova SS~Cyg, \textit{Suzaku} X-ray observations in quiescent and outburst states resulted in a maximum plasma temperature of $kT = 20.4^{+4.0}_{-2.6}$ keV and $6.0^{+0.2}_{-1.3}$ keV, with reflection amplitudes of $1.7 \pm 0.2$ and $0.9^{+0.5}_{-0.4}$, respectively, as reported by \citet{Ishida2009} and \citet{Ishida_2012}. From XMM-{\it Newton} and NuSTAR observations, \citet{2023ApJ...957...33D} showed that the maximum X-ray plasma temperature of SS~Cyg dropped from 22.8 keV in outburst to 8.4 keV in quiescence, while the reflection amplitude remained almost constant, at 1.3-1.5.

The multiple reflection scenario that we propose to explain the high reflection fraction in V648~Car (Fig.\,\ref{fig:blrefl}) requires the inner edge of its Keplerian disk to be Compton-thick and close to the WD surface ($\Delta r < h$).  This is likely to require a high accretion rate, other things being equal: the Keplerian disk will have high density, and so will the boundary layer plasma.  Since higher density plasma has a short cooling time, this will likely make the boundary layer very compact (small $\Delta r$).

When the accretion rate is modest, either the inner disk edge is far from the WD surface or the disk is optically thin, and reflection is dominated by the WD surface.  This is likely the case for SS~Cyg in quiescence even though the reflection amplitude appears greater than 1.0 in the joint \textit{NuSTAR} and \textit{XMM-Newton} observations \citep{2023ApJ...957...33D}. The recent \textit{XRISM} observation found the Fe K$\alpha$ line to be relatively narrow, excluding a significant contribution from the innermost regions of the disk -- with Keplerian velocities of several thousand km\,s$^{-1}$ \citep{Ishida2026}.

At the highest accretion rate, the boundary layer becomes optically thick and becomes a strong soft X-ray source, as is the case for SS~Cyg outburst. Note that the luminosity listed in Table\,\ref{tab:comparison} for SS~Cyg in outburst is dominated by the soft component, and the residual hard X-ray component has a luminosity of $\sim 1.0\times 10^{31}$ erg\,s$^{-1}$. The observed X-ray emission in V648~Car is luminous and hard, which in itself is a proof that its boundary layer is optically thin. The accretion rate at which the boundary layer transitions from optically thin to optically thick is not precisely known \citep{Fertig2011}, but it is expected to be a strong function of the WD mass \citep{Popham1995}. Empirically, the hard X-ray luminosity of V648~Car at $7.4 \times 10^{33}$ erg\,s$^{-1}$ implies a high accretion rate, perhaps not far below the threshold for transition to an optically thick boundary layer.

If the accretion rate is the only parameter that controls $\Delta r$, it should be the smallest in RT~Cru, followed by V648~Car, T~CrB, and finally SU~Lyn among the four symbiotic stars listed in Table\,\ref{tab:comparison}.  Thus, we can understand the relatively modest reflection amplitude in SU~Lyn and the high amplitude in V648~Car.  However, this does not explain the relatively modest value for RT~Cru.  One possibility is that this is due to a geometric effect: if our sketch (Fig.\,\ref{fig:blrefl}) is correct, we will observe more direct X-rays from the boundary layer in a face-on geometry than in an edge-on geometry.  Future observations (including a determination of the orbital inclination from optical/IR observations) would show if such an interpretation is plausible.

\section{Conclusion}

The spectral analysis of V648~Car using \textit{NuSTAR} data, consistent with what is observed from \textit{Swift}/XRT, revealed the presence of a strong Compton reflection component. We find that a significant fraction ($\sim$ 37\,\%) of the observed emission is reprocessed by cold, optically thick material near the WD. The inclusion of this component significantly improves the spectral fits and accounts for the excess observed above $\sim$ 10\,keV  (the so-called \textit{Compton hump}), which cannot be reproduced by purely thermal models. When we apply the \texttt{reflect*mkcflow} model, we confirm the high values for the reflection factor obtained ($R \sim 2$) that was also seen in previous studies \citep{Eze_2010}.

When reflection is not considered, the maximum plasma temperature is overestimated by a factor of two. As a direct consequence, the estimated $M_{\mathrm{WD}}$ is reduced by $\sim$20\,\%, from approximately $1.2\,M_{\odot}$ to about $0.95\,M_{\odot}$, under the assumption that the shock temperature reaches the maximum possible value for a Keplerian flow encountering a strong shock.  Relaxing this assumption may bring the estimated WD mass back up to $1.16\,M_{\odot}$.

Our timing analysis reveals the presence of  red noise, with no clear detection of a periodic signal.  Thus, the WD in V648~Car is likely non-magnetic. Rather, the stochastic variability observed in the light curves supports a scenario dominated by disk accretion, with the X-ray emission originating in the \textit{boundary layer}. 

We propose that the strong reflection features seen in V648~Car arise from multiple reflection of primary X-rays, originating from the boundary layer sandwiched between the WD surface and the inner edge of the Keplerian disk. Such a geometry appears plausible, and the simplified spectral model is consistent with this interpretation, but further study is necessary.  A \textit{XRISM} observation of V648~Car may test one aspect of this model: there should be a broad component to the 6.4 keV Fe K$\alpha$ line arising from reflection of the inner accretion disk, unlike the lack of such a component in SS~Cyg in quiescence \citep{Ishida2026}.

\begin{acknowledgments}

J.F. was supported by the {\it Coordenação de Aperfeiçoamento de Pessoal de Nível
Superior} (CAPES; Finance Code: 88887.992290/2024-00
and 88887.341228/2026-00). 
R.L.O. was partially supported by CNPq (PQ-315632/2023-2 and 445047/2024-0) and acknowledges financial support from NASA and the XRISM Project Science Office during his stay at NASA’s Goddard Space Flight Center.

\end{acknowledgments}

\bibliography{sample701}{}
\bibliographystyle{aasjournal}

\end{document}